\pdfoutput=1
\documentclass[%
reprint,
superscriptaddress,
nofootinbib,
 amsmath,amssymb,
 aps,
 prx,
 longbibliography,
floatfix,
]{revtex4-1}

\usepackage{graphicx}
\usepackage{dcolumn}
\usepackage{bm}
\usepackage{appendix}
\usepackage{hyperref}
\usepackage{graphicx}
\usepackage{amsmath}
\usepackage{blkarray, bigstrut}
\usepackage{amsthm}
\usepackage{algorithm}
\usepackage{algpseudocode}

\theoremstyle{definition}

\usepackage{makecell}
\usepackage{xcolor}

\usepackage{booktabs}
\usepackage{multirow}
\usepackage{mwe}

\usepackage{mathtools}

\usepackage{array,amssymb,booktabs}
\newcolumntype{C}{>{$}c<{$}}
\AtBeginDocument{
    \heavyrulewidth=.08em
    \lightrulewidth=.05em
    \cmidrulewidth=.03em
    \belowrulesep=.65ex
    \belowbottomsep=0pt
    \aboverulesep=.4ex
    \abovetopsep=0pt
    \cmidrulesep=\doublerulesep
    \cmidrulekern=.5em
    \defaultaddspace=.5em
}
\usepackage[utf8]{inputenc}

\usepackage{color}
\definecolor{keywordsColor}{rgb}{0,0,0.5}
\definecolor{commentsColor}{rgb}{0.6,0,0}
\definecolor{stringColor}{rgb}{0,0.5,0}
\definecolor{gray}{rgb}{0.9,0.9,0.9}

\DeclareFixedFont{\ttb}{T1}{txtt}{bx}{n}{6} 
\DeclareFixedFont{\ttm}{T1}{txtt}{m}{n}{6}  
\DeclareFixedFont{\ttn}{T1}{txtt}{m}{n}{10}  

\usepackage{listings}
\newcommand\pythonstyle{\lstset{
backgroundcolor=\color{gray},   
  basicstyle=\scriptsize\ttfamily\ttm,        
  breakatwhitespace=false,         
  breaklines=true,                 
  captionpos=b,                    
  commentstyle=\color{commentsColor}\textit\ttfamily\ttm,    
  frame=tb,	                   	   
  keepspaces=true,                 
  keywordstyle=\color{keywordsColor}\bfseries\ttfamily\ttb,       
  language=Python,                 
  deletekeywords={Abs, Int, ...},            
  rulecolor=\color{black},         
  showspaces=false,                
  showstringspaces=false,          
  showtabs=false,                  
  stepnumber=1,                    
  stringstyle=\color{stringColor}\ttfamily\ttm, 
  tabsize=2,	                   
  title=\lstname,                  
  columns=fixed                    
}}

\newcommand{\zh}[1]{\textcolor{blue}{[Zhanghao: #1]}}

\lstnewenvironment{python}[1][]
{
\pythonstyle
\lstset{#1}
}
{}

\begin{document}

\preprint{APS/123-QED}

\title{Scalable Low-latency Optical Phase Sensor Array}

\author{Zhanghao Sun$^1$, Sunil Pai$^1$, Carson Valdez$^1$, Maziyar Milanizadeh$^2$, Andrea Melloni$^2$, Francesco Morichetti$^2$, David A. B. Miller$^1$, Olav Solgaard}
\email{zhsun@stanford.edu}
\address{Ginzton Laboratory, Stanford University, 348 Via Pueblo Mall, Stanford, CA 94305, USA\\
$^2$Department of Electronics, Information and Bioengineering, Politecnico di Milano, via Ponzio 34/5, 20133, Milano, Italy}

\begin{abstract}

    Optical phase measurement is critical for many applications and traditional approaches often suffer from mechanical instability, temporal latency, and computational complexity. In this paper, we describe compact phase sensor arrays based on integrated photonics, which enable accurate and scalable reference-free phase sensing in a few measurement steps. This is achieved by connecting multiple two-port phase sensors into a graph to measure relative phases between neighboring and distant spatial locations. We propose an efficient post-processing algorithm, as well as circuit design rules to reduce random and biased error accumulations. We demonstrate the effectiveness of our system in both simulations and experiments with photonic integrated circuits. The proposed system measures the optical phase directly without the need for external references or spatial light modulators, thus providing significant benefits for applications including microscope imaging and optical phased arrays. 

\end{abstract}

\maketitle

\section{Introduction}

Optical phase measurements are important in imaging~\cite{microscopy1, microscopy2, 3d1, opa_lidar1, opa_lidar2}, environmental sensing~\cite{env_sense_1, env_sense_2}, optical communications~\cite{dab_communication, joe_communication, opa_comm1, opa_comm2}, and optical neural networks~\cite{sunil1, sunil2, onn1, onn2}. Previous approaches for optical phase measurement can be divided into two categories: reference-based and reference-free. The first approach is based on homodyne/heterodyne interference between the detected optical field and a reference signal directly routed from the light source~\cite{homo1, hetero1}. 
Such systems benefit from the improved detection signal-noise-ratio (SNR) while suffering from any instabilities in the reference arm~\cite{hetero1, hetero4, hetero_calib2, hetero_calib1}. 
The reference-free approach avoids the need for an external reference by interfering different components of the incident field with one another~\cite{self-ref1, self-ref2, wish}. Phase contrast microscopy exploits a phase shifting aperture to interfere the incident light field with a plane wave and acquires an approximate phase profile for transparent samples~\cite{microscopy1}. Modern reference-free wavefront measurements typically use a programmable spatial light modulator (SLM) together with post-processing with phase retrieval algorithms that lead to significant computational cost~\cite{wish, slm_ws1, slm_ws2}. In a special case, the Shack-Hartmann (SH) wavefront sensor~\cite{sh_sensor} employs a micro-lens array instead of an SLM. 





Recently, a reference-free optical field measurement based on photonics integrated circuit (PIC) has been proposed and demonstrated~\cite{dab_selfconfig, ref_free_exp}. 
During operation, Mach-Zehnder interferometers (MZIs) in the PIC are \textit{progressively} configured through power minimizations, and the relative input optical phase over all the inputs is calculated based on the phase settings. This approach is robust against vibrations~\cite{int_hetero1}, but it requires that the MZIs are controlled by analog voltages of high precision, and its progressive nature of successive minimizations introduces significant time delay.

In this paper, we propose a compact photonic phase sensor array for reference-free phase sensing requiring only a few measurements. This is achieved by connecting multiple two-port phase sensors into a graph to measure relative phases (phase gradients) between both neighboring and \textit{distant} spatial locations. Thanks to the versatile PIC platform, all measurements are conducted in parallel in a single photonics focal-plane~\cite{3d1}. 
Compared to the progressive PIC phase measurement, our system has lower latency and requires less hardware.

To minimize measurement errors caused by sensor noise, we add post-processing, formulated as a least-square problem. Compared with non-convex phase retrieval algorithms, the post-processing in our system is fast and robust. 
We analyze the phase sensor array's robustness to noise and hardware errors for large numbers of input ports. Phase sensing accuracy is shown to be determined by the connectivity of the photonics circuit. Errors can be significantly reduced by (1) introducing distant relative phase measurements, (2) appropriately placing phase shifters in phase sensors, and (3) introducing redundant relative phase measurements.
We demonstrate the effectiveness of our designs both with simulations and through experiments. The proposed system is promising in various applications, including phase imaging in microscopy and in optical phased arrays (OPA).~\cite{opa_review}.

\section{Integrated phase sensor} 
\label{sec:single_sensor}

\begin{figure}
    \centering
    \includegraphics[width=1.0\linewidth]{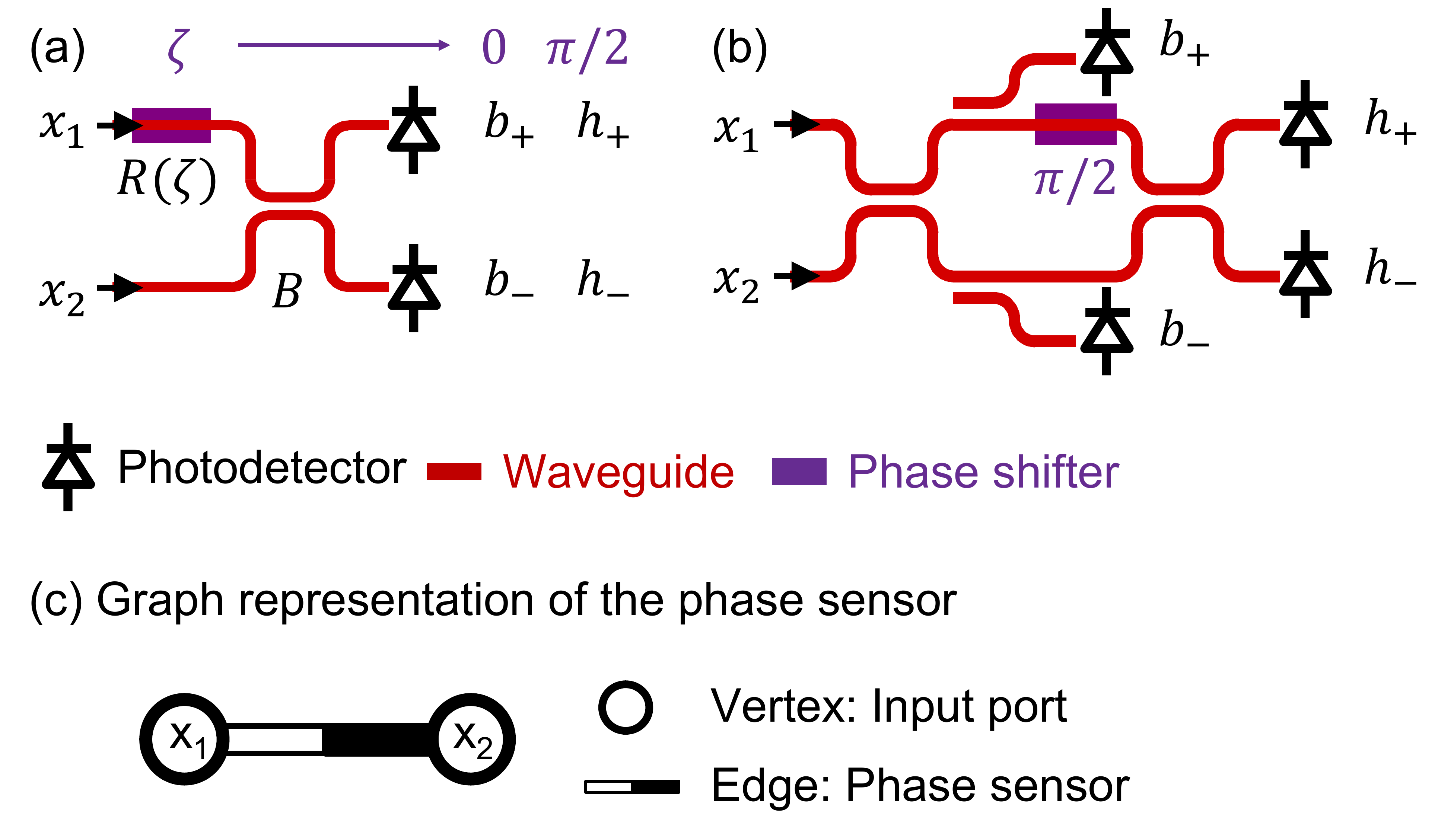}
    \caption{(a) Active phase sensor using a phase shifter followed by a beamsplitter. (b) Passive phase sensor. (c) Graph representation of a single phase sensor. The vertex (circle) represents an input port, while the edge represents a phase sensor. The polarity of the edge represents the placement of the phase shifter in the sensor. Formally, the white end of the edge in (c) corresponds to the port or side of the phase sensor that has the phase shifter, as in (a), or the $\pi/2$ phase shift, as in (b).}
    \label{fig:single_sensor}
\end{figure}

We start with the description of a single two-port phase sensor, which detects the relative phases between two input optical fields $x_1$, $x_2$ that enter the phase sensor in single mode waveguides. Similar device have recently been applied in reference-based detection systems~\cite{int_hetero1, 3d1}. As shown in Fig.~\ref{fig:single_sensor} (a), a straight-forward implementation uses in-phase and quardrature (I/Q) interferometric detection~\cite{int_hetero1} with a $50/50$ beam splitter and a tunable phase shifter. The system can be modeled as a $2$-in, $2$-out linear system with transmission matrix $H$:
\begin{flalign}
\label{eqn:mzi}
H = BR(\zeta) = \frac{1}{\sqrt{2}}\begin{bmatrix}1 & i \\ i & 1 \end{bmatrix} \begin{bmatrix} e^{i\zeta} & 0 \\ 0 & 1 \end{bmatrix}
\end{flalign}

Here $B$, $R$ are the transmission matrices of the $50/50$ beam splitter and phase shifter. When the phase shift $\zeta$ is set to $0$ ($b_+$ and $b_-$ outputs) and $\pi/2$ ($h_+$ and $h_-$), the detected optical powers in the two photodetectors are: 
\begin{flalign}
\label{eqn:single}
&h_\pm = (|x_1|^2 + |x_2|^2 \pm 2|x_1||x_2|\textrm{cos}\gamma)/2,
\\
\nonumber
&b_\pm = (|x_1|^2 + |x_2|^2 \pm 2|x_1||x_2|\textrm{sin}\gamma)/2
\end{flalign}

where $\gamma$ is the relative phase between $x_1$, $x_2$. Eqn.~\ref{eqn:single} gives the estimate $\gamma = \textrm{arctan}((b_+ - b_-)/(h_+ - h_-))$. This active sensor design requires two measurement steps and utilizes one phase shifter whose phase delay is switched between $0$ and $\pi/2$. To avoid the delay of actively tuning the phase shifter, we use the passive sensor shown in Fig.~\ref{fig:single_sensor} (b), in which $h_\pm$ and $b_\pm$ are measured in one step, although the optical power in each photodetector is halved (We assume all directional couplers in the circuit are $50/50$ beam splitters, while the two on the sides can be designed to have a same beam splitting ratio other than $50/50$). An advantage of the passive phase sensor is that it has a smaller footprint\footnote{In Fig.~\ref{fig:single_sensor}, the devices are not to scale.} because we avoid the long active phase shifter, and no control circuitry is required.

To simplify the following discussions and visualizations, we abstract the phase sensor into a graph without the physical implementation details\footnote{For review on graph concepts, please refer to ~\cite{graph_textbook}}. As shown in Fig.~\ref{fig:single_sensor} (c), each input port is represented by a vertex in the graph, and each phase sensor is represented by an edge. Since the phase shifter is only applied to one arm in the sensor, the phase sensor, and hence the edge that represents it, is asymmetrical, and we define the input port on the side with the phase shifter or on the side with the $\pi/2$ phase shift path to have positive polarity (the white end of the white-black edge)\footnote{A more standard concept in graph theory is a directed edge. However, we use polarity to represent this asymmetry to avoid confusion with the optical power propagation direction.}. 




\begin{figure}
    \centering
    \includegraphics[width=0.9\linewidth]{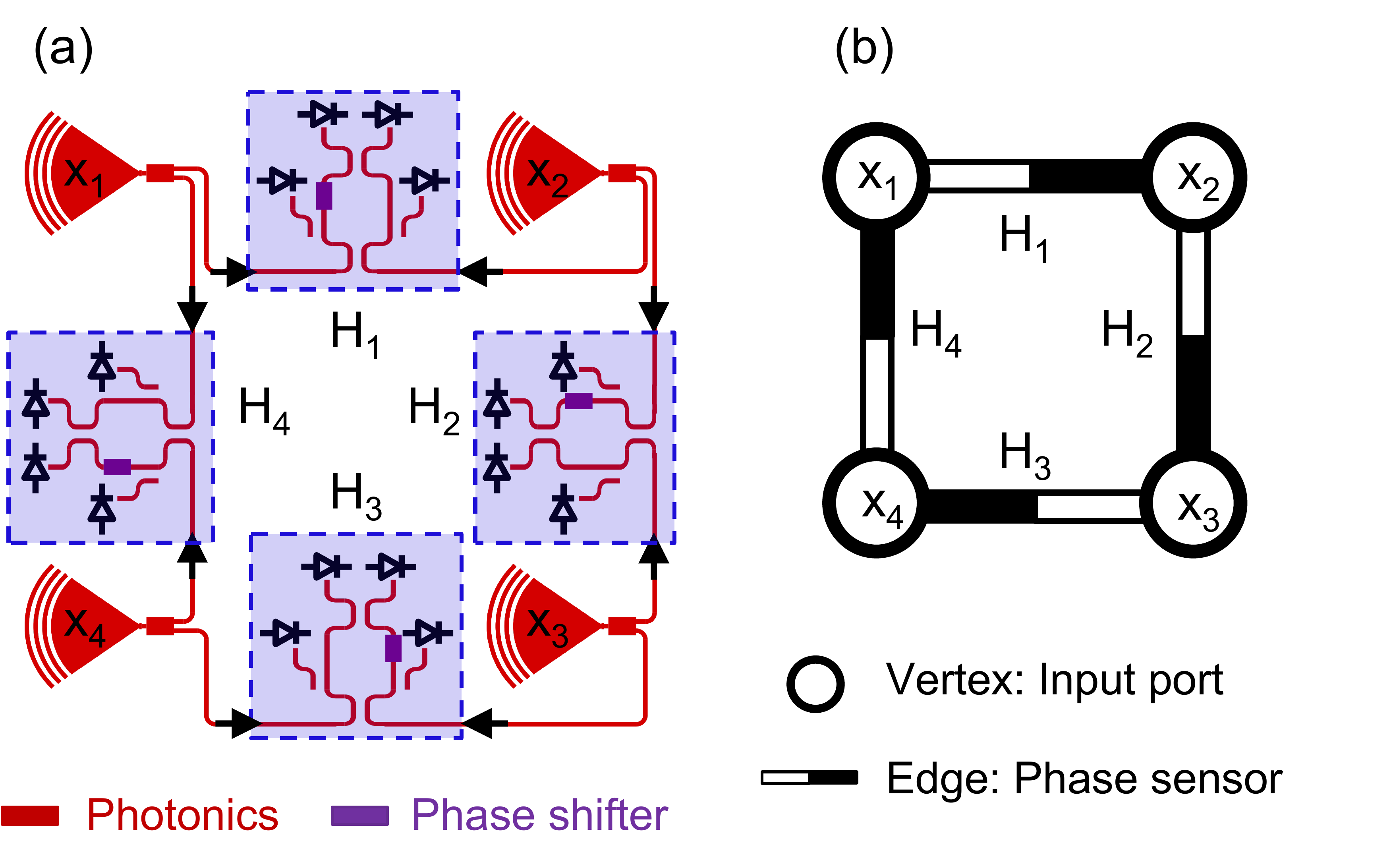}
    \caption{(a) Simple example of phase sensor array with 4 input ports and 4 phase sensors (not to scale) and (b) corresponding  graph representation. Black arrows in (a) show the optical power propagation directions and black/white colored edges in (b) represent the phase sensors with polarities. To handle the measurement errors, a least-square method is used to compute the phase profile from phase sensor measurements.}
    \label{fig:least_square}
\end{figure}

\section{Phase sensor array}
\label{sec:phase_sensor_array}
Multiple phase sensors can be integrated into an array to measure the phase profile across multiple input ports, as shown in Fig.~\ref{fig:least_square} (a). 
When the system is used to measure a free-space light field, an optical interface converts free-space incident light to the phase sensor using grating couplers, photonic lantern, or other free-space optical interface.~\cite{grating_coupler1, grating_coupler2, photonic_lantern} (please refer to supplementary material for more discussions).
At each input port of the sensor, the received optical signal is split into multiple single-mode waveguides~\cite{splitter} and then fed into the phase sensors, as shown by the black arrows in Fig.~\ref{fig:least_square} (a). Following the conventions defined in Sec.~\ref{sec:single_sensor}, we also show the corresponding graph representation in Fig.~\ref{fig:least_square} (b). Each phase sensor is accessed independently, and the input light is routed to the photodetectors and converted into electronic signals. The electronic readout circuitry can be similar to that of common image sensors~\cite{cmos_camera}.


From the readout of photodetectors, the phase measurement of each phase sensor is calculated using Eqn.~\ref{eqn:single}. 
For the example shown in Fig.~\ref{fig:least_square}, the sensor array has $4$ input ports $\{x_1, x_2, x_3, x_4\}$ (without loss of generality, we select $x_1$ to be $0$ phase). We need to find $3$ unknown phase differences relative to the phase of $x_1$, $\mathbf{p} = \{p_2, p_3, p_4\}$ and there are $4$ phase sensors $\{H_1, H_2, H_3, H_4\}$ in the array with measurement results $\mathbf{p}_H = \{p_{H_1}, p_{H_2}, p_{H_3}, p_{H_4}\}$. The measurement process can be expressed in a matrix $M$

\begin{eqnarray}
\label{eqn:measure}
\nonumber
\mathbf{p}_H  \;\;\;\;\;\;\;\;\;\;\;\;\;\;\;\;\;\;\;\;  M \;\;\;\;\;\;\;\;\;\;\;\;\;\;\;\;\; \mathbf{p} \;\;\;  \;\;\;\;\;\;\; \mathbf{e}_H 
\;\;\;
\\
\left[
  \begin{matrix}
    p_{H_1}  \\[-5pt] p_{H_2} \\[-5pt] p_{H_3} \\[-5pt] p_{H_4}
  \end{matrix}
\right] = 
\left[
  \begin{matrix}
    1 & -1 & 0 & 0  \\[-5pt] 0 & 1 & -1 & 0 \\[-5pt] 0 & 0 & 1 & -1 \\[-5pt] -1 & 0 & 0 & 1 
  \end{matrix}
\right]
\left[
  \begin{matrix}
    0  \\[-5pt] p_{2} \\[-5pt] p_{3} \\[-5pt] p_{4}
  \end{matrix}
\right]
+ 
\left[
  \begin{matrix}
    e_{H_1}  \\[-5pt] e_{H_2} \\[-5pt] e_{H_3} \\[-5pt] e_{H_4}
  \end{matrix}
\right]
\end{eqnarray}

where $\mathbf{e}_H = \{e_{H_i}\}$ are the phase measurement errors. 
When the number of phase sensors is one less than the number of input ports (we call it a ``non-redundant'' array in Sec.~\ref{sec:non-redudant}), $\mathbf{p}$ can be recovered from $\mathbf{p}_H$ by algebraic calculations. However, in more general cases (e.g., Fig.~\ref{fig:least_square}), number of phase measurements exceeds the number of unknown phases. In those cases we solve a least-square problem to better account for the measurement errors, with the pseudo-inverse of the measurement matrix $M^{+} = (M^{T}M)^{-1}M^{T}$~\cite{linalg_textbook}.


\begin{align}
\label{eqn:least_square}
&\left\{ \begin{array}{l}
\hat{\textbf{p}} = \textrm{argmin}_{\textbf{p}} |\textbf{p}_H - M\textbf{p}|^2 = M^+\textbf{p}_{H}\\
\textbf{p} = (0, p_{2}, p_{3}, p_{4})^T \\
\textbf{p}_H = (p_{H_1}, p_{H_2}, p_{H_3}, p_{H_4})^T
\end{array} 
\right.
\end{align}

where $\hat{\textbf{p}}$ is the estimated phase profile. In our system, each row in $M$ corresponds to the measurement made by one phase sensor. The measurement matrix has a $1$ element and an $-1$ element in each row, with all other elements $0$. Switching the position of the phase shifter from one arm to the other arm in the phase sensor results in a change of sign in the corresponding row in matrix $M$. Since matrix $M$ is determined once the device is fabricated, so is the pseudo-inverse $M^+$ (which can be deduced from the singular value decomposition of $M$). Solving this least-square problem then only requires a simple matrix-vector multiplication between $M^{+}$ and the vector of measurements $\mathbf{p}_H$.

Note that due to the $2\pi$ ambiguity of phase, the first line of Eqn.~\ref{eqn:least_square} should be expressed as $\hat{\textbf{p}} = \textrm{argmin}_{\textbf{p}} |\textrm{mod}[(\textbf{p}_H - M\textbf{p}), 2\pi]|^2$, where $\textrm{mod}[\;, 2\pi]$ is the modulo operation. This is different from a normal least square problem. However, a simple workaround converts it into a normal least square problem (please refer to supplementary material for more details). Therefore, we still use the normal least square optimization formula for simplicity.

\section{Phase Sensing Error}
\label{sec:phase_error}

There are two major sources of error in the phase measurements. The first is noise in photodetection, including photon shot noise~\cite{noise1, noise2}, dark current~\cite{dark_current}, and amplifier readout noise~\cite{readout}. 
The second source is systematic hardware errors, including incorrect beam splitting ratios and phase shifts~\cite{hardware_error, dab_imperfect}, and non-uniform photodetector gain~\cite{hetero_calib1}. Such systematic errors can be partially removed by calibration (please refer to supplementary material for more details on the calibration process). Typical residual errors in beam-splitting ratio reported in the literature are $\sim 1\%$ to $10\%$~\cite{hardware_error, sunil1}. Note that hardware errors can be random or biased. Random hardware error means every single element in the system has different imperfections~\cite{imperfect_1, imperfect_2}, while biased hardware error means all elements in the system are biased to the same erroneous state. Such biased error can be due to temperature, optical wavelength, fabrication, or other global factors. In the following, we only discuss beam-splitting errors, while all the conclusions can be straight-forwardly applied to other types of hardware errors (please refer to supplementary material for details).

We use simulations to evaluate the influence of noise and hardware errors on phase sensing accuracy. We first consider a single-phase sensor. Under specific noise and hardware errors, we use the expectation of absolute error of measured relative phase $\mathbb{E}[\textrm{RMS}] = \mathbb{E}[||\hat{p} - p||]$ to characterize the average phase sensing error. Following the convention in~\cite{hardware_error}, the expectation is taken over a set of random input fields.

\begin{eqnarray}
\label{eqn:input}
&x \sim (\mathcal{N}(0,1) + i\mathcal{N}(0,1))/\sqrt{2},\;\; \mathbb{E}[|x|^2] = 1
\\
\nonumber
&h'_\pm = h_\pm + \sigma_{n}\mathcal{N}(0,1),\;\; b'_\pm = b_\pm + \sigma_{n}\mathcal{N}(0,1)
\\
\nonumber
&\mathrm{SNR} := \mathbb{E}[|x|^2]/\sigma_{n} = 1/\sigma_n
\end{eqnarray}
where $\mathcal{N}(0,1)$ is the normal distribution with mean $0$ and standard deviation $1$, $h'_\pm$ and $b'_\pm$ are the noisy measurements and $\sigma_{n}$ is the magnitude of measurement noise. For simplicity, we use a simple noise model and define the overall signal-noise-ratio (SNR) of the system as $1/\sigma_n$.
For hardware errors, we simulate random and biased beam splitter errors (BSE). Similar to Eqn.~\ref{eqn:mzi}, the imperfect beam splitter with an uneven splitting ratio is modeled by the $2\times2$ matrix $B' = \frac{1}{\sqrt{2}}\begin{bmatrix}\sqrt{1+\epsilon} & i\sqrt{1-\epsilon} \\ i\sqrt{1-\epsilon} & \sqrt{1+\epsilon} \end{bmatrix}$, where $\epsilon$ is the error. For the random error case, we assume $\epsilon \sim \sigma_\epsilon \; \mathcal{N}(0,1)$. For the biased error case, we use the same $\epsilon$ for all beam splitters. For all hardware error simulations, we still include the photodetection noise.

\begin{figure}
    \centering
    \includegraphics[width=1.0\linewidth]{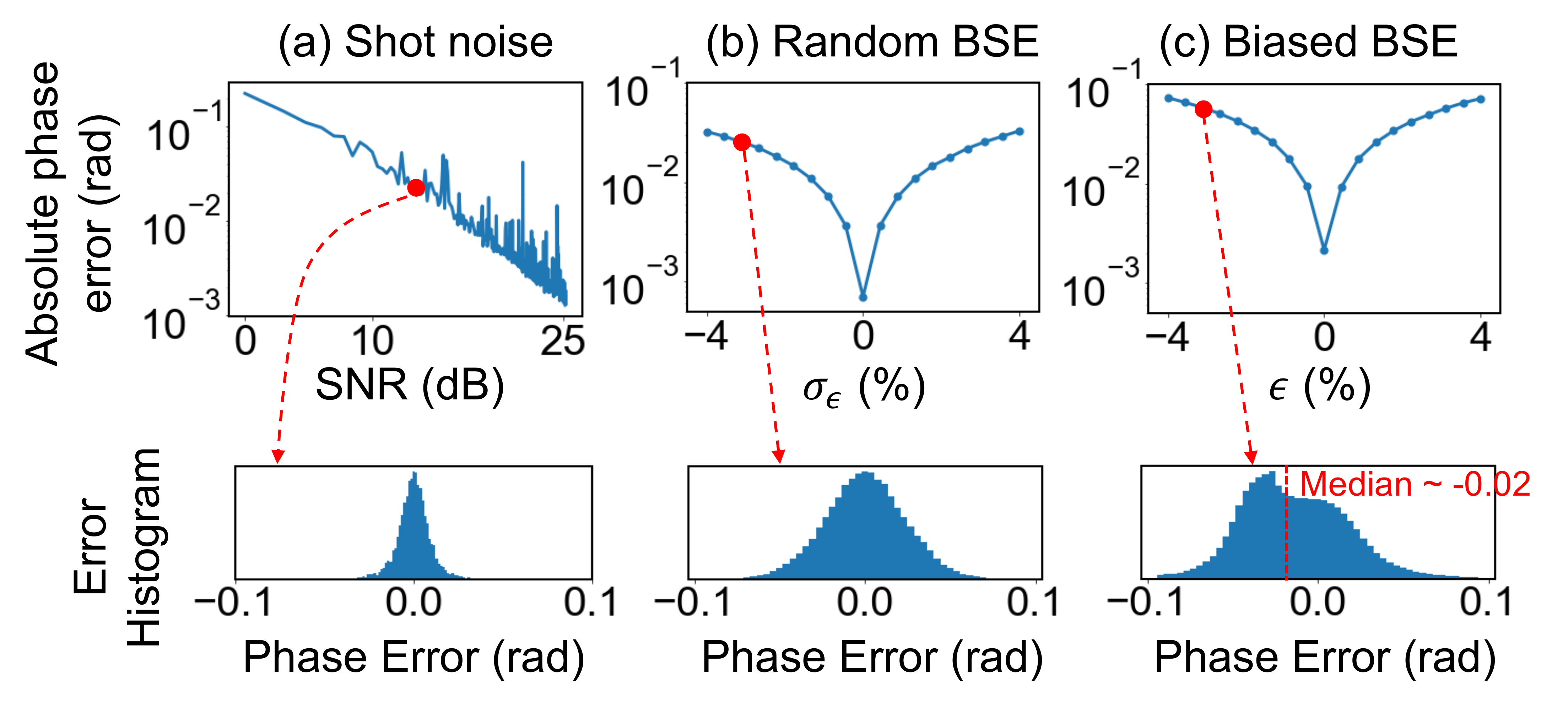}
    \caption{Single sensor phase measurement error (absolute values averaged over 1000 random input fields simulations) with (a) photodetection noise, (b) random beam splitter error, and (c) biased beam splitter error. Second row: corresponding measurement error histograms at the red dots indicated settings (over 1000 random input fields simulations)).}
    \label{fig:single_sensor_error}
\end{figure}

The simulated average phase sensing errors in Fig.~\ref{fig:single_sensor_error} (a)-(c) show several important results. First, except under extremely low SNR, average phase sensing error is inversely proportional to the SNR when no hardware error is present. Second, hardware errors significantly increase phase sensing errors. Third, biased hardware errors cause biased phase sensing error (as shown in Fig.~\ref{fig:single_sensor_error}(c), with median $\sim -0.02$).


\section{Phase Sensing Error in Array}
\label{sec:scalability}

We denote the phase sensing error at phase sensor $i$ as $e_{Hi}$. Following the simulation results in Fig.~\ref{fig:single_sensor_error}, we can decompose the statistical distributions of phase sensing error into a standard deviation $\sigma_e$ and a bias $e_0$. Since the phase profile in a sensor array is estimated by Eqn.~\ref{eqn:least_square}, we characterize the average \textit{estimated} phase sensing accuracy with the expected root mean square error (RMSE) $\mathbb{E}[\textrm{RMS}] = \mathbb{E}[||\hat{\textbf{p}} - \textbf{p}||]$. Similar to Sec.~\ref{sec:phase_error},  the expected value is taken over random input fields following Eqn.~\ref{eqn:input}. 

\begin{figure}
    \centering
    \includegraphics[width=0.9\linewidth]{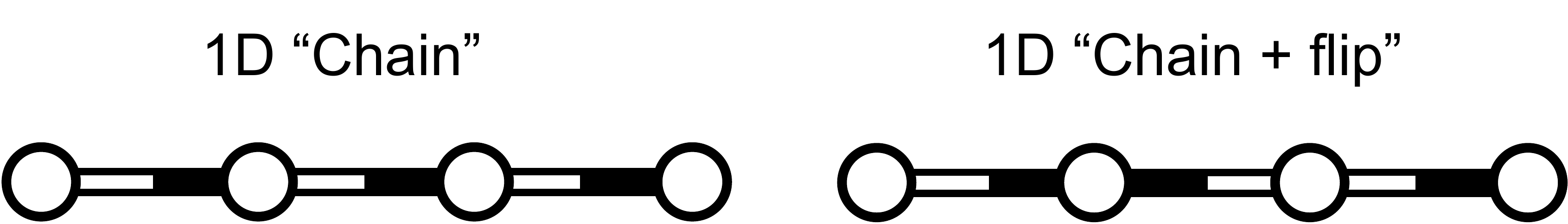}
    \caption{Left: ``1D chain'' configuration. Right: ``1D chain + flip'' configuration. While random error accumulation maintains the same, biased error accumulation is eliminated.}
    \label{fig:error_example}
\end{figure}

Fig.~\ref{fig:error_example} left side (``1D chain'' configuration) shows an example of the phase sensor array. 
The expectation of phase RMSE for a phase sensor array with $N$ input ports can be expressed as

\begin{align}
\label{eqn:chain_error}
&\mathbb{E}[(\hat{p}_i - p_i)^2] = \sum_{j=1}^{i} \mathbb{E}[e_{Hi}^2] = i\sigma_e^2 + i^2e_0^2
\\
\nonumber
&\mathbb{E}[\textrm{RMS}] = \sqrt{\mathbb{E}[\frac{1}{N}\sum_{i=1}^{N-1} (\hat{p}_i - p_i)^2]}
\\
\nonumber
&= \sqrt{\sigma_e^2(N-1)/2 + e_0^2(N+1)(2N+1)/6}
\\
\nonumber
&= \sqrt{\textrm{E}_r^2 + \textrm{E}_b^2}
\end{align}

We can divide $\mathbb{E}[\textrm{RMS}]$ into contributions from the random error in each sensor $\sigma_e$ ($\textrm{E}_r$, scales with $O(\sqrt{N})$) and the bias error $e_0$ ($\textrm{E}_b$, scales with $O(N)$). This error becomes intolerable with moderate $N$ thus prohibiting scaling up. 

Eqn.~\ref{eqn:chain_error} also indicates that the expected error at the $i$th port ($\mathbb{E}[(\hat{p}_i - p_i)^2]$) depends on the number of intermediate connections (path length in graph theory) between it and the reference port, as shown in Fig.~\ref{fig:error_example}. The shorter the path length, the less accumulated errors. The key difference between the random and biased errors lies in their dependencies on the polarities of the phase sensors. $\textrm{E}_r$ does not change when the polarities of the connections change, while $\textrm{E}_b$ largely depends on the polarities. By flipping the edge (phase sensor) polarities in the 1D chain configuration, $\textrm{E}_b = e_0$ (constant w.r.t $N$) is achieved in ``1D chain + flip'' configuration (Fig.~\ref{fig:error_example} right side).

\begin{figure*}
    \centering
    \includegraphics[width=1.0\linewidth]{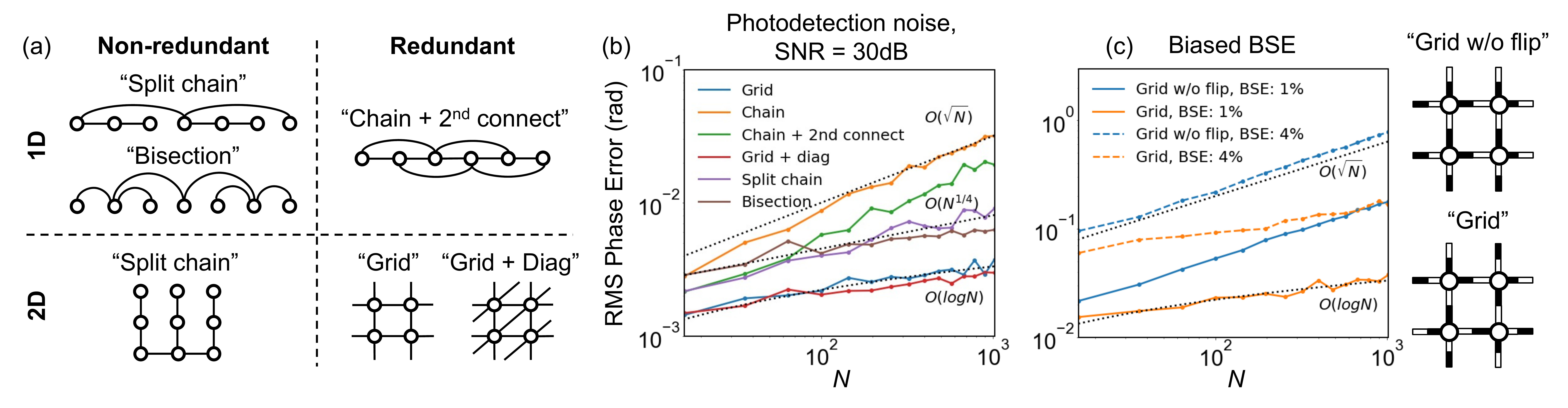}
    \caption{(a) Configurations categorized into 1D, 2D, non-redundant, and redundant. (b) RMS phase sensing error with respect to the number of input ports $N$ when photodetection noise is present ($\textrm{SNR} = 30$dB). Designs following the design rule (e.g., ``split chain'', ``grid'') achieve better scalability compared to the naive chain configuration. (c) Impact of phase sensor polarity selection with biased error. Grid configuration achieves better scalability by flipping the edge polarities in ``Grid w/o flip'' configuration.}
    \label{fig:configs}
\end{figure*}

For more complex circuit configurations, analytical relationships between $N$ and $\textrm{E}_r$, $\textrm{E}_b$ are difficult to obtain. Nonetheless, following Eqn.~\ref{eqn:least_square}, we can establish a simple relationship between the RMSE and $M^+$, the pseudo-inverse of measurement matrix $M$ (derivations are presented in supplementary material)

\begin{eqnarray}
\label{eqn:array_error}
\left\{ \begin{array}{l}
\textrm{E}_r = \sigma_e\sqrt{\sum_{ij}M^+[ij]^2}
\\
\textrm{E}_b = e_0\sqrt{\sum_{ij}(M^{+T}M^+)[ij]}
\end{array} 
\right.
\\
\nonumber
M^+ = (M^TM)^{-1}M^T
\end{eqnarray}

Note that $\sqrt{\sum_{ij}M^+[ij]^2}$ and $\sqrt{\sum_{ij}(M^{+T}M^+)[ij]}$ are the Frobenius and $L_{2,1}$ norms of matrix $M^+$, and are related to ``graph spectrum'' in graph theory~\cite{graph_spectral_textbook}. From Eqn.~\ref{eqn:array_error}, we again observe that flipping signs of the rows in matrix $M$ will not change $\textrm{E}_r$, but has an influence on $\textrm{E}_b$. Therefore, these two metrics are decoupled and can be considered separately in the circuit configuration design: We first ignore the polarities in the circuit graph and minimize $\textrm{E}_r$ by placing edges (phase sensors) at proper positions. In particular we want to minimize the path lengths from the reference port to other ports in the circuit graph. Then we flip the polarities for part of the edges to minimize $\textrm{E}_b$. Due to this decoupled design rule, in the following, we ignore the phase sensor polarity when discussing $\textrm{E}_r$ and we always assume that there exists a phase sensor polarity arrangement to minimize $\textrm{E}_b$.


\subsection{Non-redundant array}
\label{sec:non-redudant}
The minimum number of phase sensors needed to measure full phase profile with $N$ input ports is $N-1$. We denote this type of circuits as non-redundant. Several examples are shown in Fig.~\ref{fig:configs} (a) (polarities of phase sensors are ignored, as discussed above). Since only one path exists between the phase reference input to any other input, the design rule of ``shorter path, lower error'' holds. The most naive chain configuration has $\textrm{E}_r \sim O(\sqrt{N})$, which becomes intolerable with moderate $N$. By breaking the long path into shorter ones, $\textrm{E}_r$ is reduced to $O(N^{1/4})$ in ``split chain'' configuration. With $N = 1000$, this is $\sim 6\times$ smaller than the simple 1D chain configuration of Fig.~\ref{fig:error_example}. Similar configurations can also be implemented with input ports arranged in 2D. Through bi-sectioning the long path, the ``bi-section'' configuration achieves $O(\sqrt{\textrm{log}N})$ RMS error. However, this configuration requires complicated waveguide routings that increases the footprint significantly. In all these configurations, $\textrm{E}_b$ can be reduced to $O(1)$ (constant) by flipping the polarities of edges.

We simulate the average phase sensing accuracy with photodetection noise (Fig.~\ref{fig:configs} (b)) $\textrm{SNR} = 30$dB. The simulation results fits well with the theoretical analysis. 
 
\subsection{Redundant array}
\label{sec:redundant}
Redundant phase sensing elements can be introduced into the circuit as shown in Fig.~\ref{fig:configs} (a). In the redundant array, multiple paths exist between the phase reference input and other unknown phase inputs, so the least square optimization is effectively performing an averaging over phase estimation results accumulated along all paths. With more redundancies, more paths are involved in the averaging and thus provides a more accurate estimate. On the other hand, with more redundancies, each input port is connected to more phase sensors, and each phase sensor is receiving less optical power, resulting in lower signal-noise-ratio.

We show two representative cases in Fig.~\ref{fig:configs} (b). In the best case, $O(\textrm{log}N)$ is achieved in the grid configuration with flipped edges. The ``grid + diag'' configuration has $50\%$ more edges (phase sensors) compared to the grid configuration. However, the phase sensing accuracies are similar, due to the trade-off discussed above. We also show an example with biased beam splitter errors (Fig.~\ref{fig:configs} (c)). Similar to single phase sensor simulations, we use biased BSE $= 1\%$, $4\%$ in this simulation. It can be seen that with correctly flipped edges, phase sensing error maintains $O(\textrm{log}N)$ scalability, while failing to do so results in phase sensing error $\sim O(\sqrt{N})$.

\begin{figure*}[t]
    \includegraphics[width=1.0\linewidth]{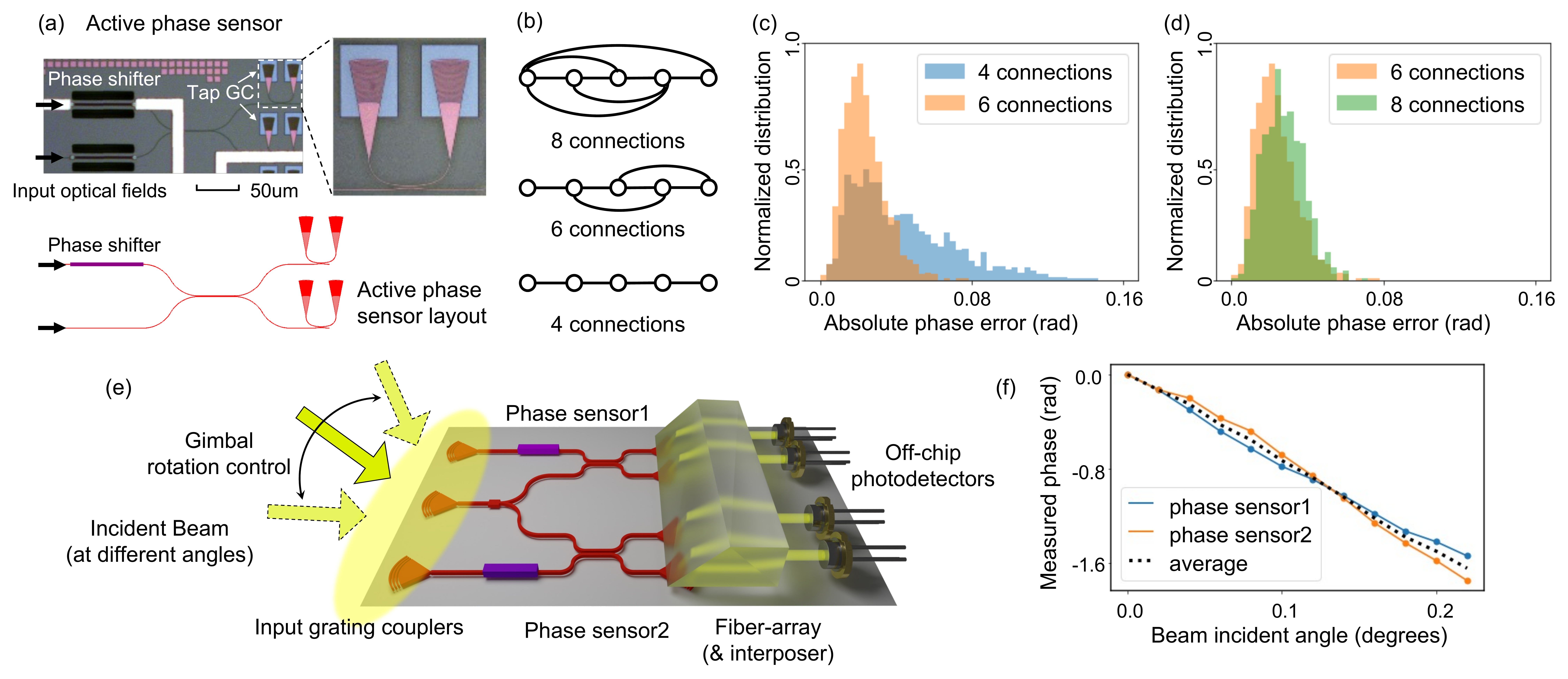}
    \caption{Experimental demonstration of redundant phase sensor array with $5$ input ports. (a) Fabricated active phase sensor on Silicon PIC with thermal phase shifter and tap grating coupler (tap GC)~\cite{sunil_insitu}. (b) Non-redundant circuit configuration with $4$ phase sensors and redundant configuration with $6$ phase sensors. (c),(d) Absolute phase sensing error distributions over $1000$ experiments with random input light field. 6 connections configuration reduces error by $\sim 2\times$ compared to non-redundant 4 connections configuration, while 8 connections configuration leads to slightly worse accuracy due to the trade-off between system redundancy and per-sensor SNR with a small number of ports. (e),(f) Experimentally measuring the phase of an incident plane wave at three input ports (two phase sensors). We use a gimbal to control incident angle of the beam. Our setup presents a good linear relationship between the relative phase measurement and the beam incident angle. }
 \label{fig:real_exp}
\end{figure*}

\begin{figure*}
    \centering
    \includegraphics[width=1.0\linewidth]{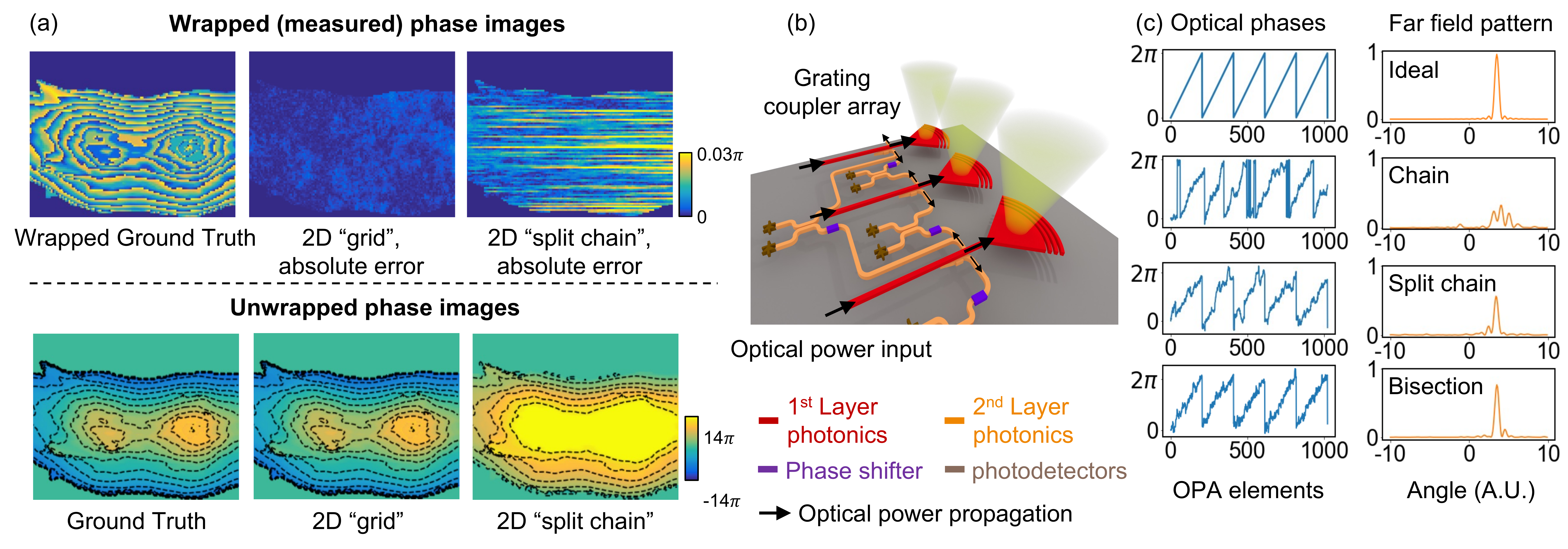}
    \caption{(a) Comparisons on $128\times128$ resolution phase imaging and unwrapping with photodetection noise. The redundant 2D grid sensor array achieves more robust performance. (b) Schematic showing the application of proposed phase sensing array in optical phased array (OPA) phase monitoring. A dual-layer photonics circuit is used for simpler routing. Small amount of power is tapped out from the first layer into the second layer for phase monitoring. We use black arrows to visualize the optical power application. (c) Optical phase profile and far field pattern in $N = 1024$ 1D OPA, with ``chain'', ``split chain'', and ``bisection'' phase sensor arrays for phase monitoring. With better phase monitoring design, phase error is significantly reduced and far field pattern has better side-lobe suppression.}
    \label{fig:application}
\end{figure*}

\subsection{Experimental results}
We demonstrate the effectiveness of redundant phase sensor arrays on a fabricated Silicon PIC platform (shown in Fig.~\ref{fig:real_exp} (a))~\cite{sunil_insitu}. We use an on-chip circuit to detect optical power, and emulate different connection configurations between phase sensors (refer to supplementary for detailed hardware implementation). We use tap grating couplers (tap GC) as an alternative for integrated photodetectors, following the design in ~\cite{sunil_insitu}. The tap GC couples out $3\%$ optical power from the waveguide and emits to an IR camera sitting on top of the device. Our circuit consists of $5$ input ports, and the non-redundant configuration consists of $4$ phase sensors (Fig.~\ref{fig:real_exp} (b) upper), while the redundant phase sensor array utilizes $6$ phase sensors (Fig.~\ref{fig:real_exp} (b) lower). We generate $1000$ random input light fields following Eqn.~\ref{eqn:input}, and measure the phase profile following Eqn.~\ref{eqn:least_square}. Despite the optical power in each phase sensor being lower, the redundancy reduces the average phase sensing error by $\sim 2\times$. As shown in Fig.~\ref{fig:real_exp} (c), the redundant circuit design is beneficial in this proof-of-concept demonstration, consistent with the simulation results discussed in Sec.~\ref{sec:redundant}. As the PIC scales, the benefit from redundant circuit configuration is expected to be more significant.

Note that excessive amount of redundancy can also be unfavorable. As shown in Fig.~\ref{fig:real_exp} (b)(d), we tested another configuration with 8 connections (more redundant relative phase measurements). However, the 8 connections configuration achieves slightly worse accuracy compared to the 6 connections configuration. This is due to the trade-off between system redundancy and per-sensor SNR. With excessive redundancies the input optical power is too distributed and SNR in each phase sensor is low. With a small number of input ports, this effect overwhelms the benefit from robustness in phase profile estimation process, thus leads to less favorable performance.  

In Fig.~\ref{fig:real_exp} (e),(f), we further demonstrate measuring the phase profile of free-space incident light with the PIC phase sensor array. We shine a plane wave onto three input ports (two phase sensors) and use a gimbal to control the angle of the incident beam. We use a fiber array to couple light off the chip and measure the optical power with off-chip photodetectors to avoid stray lights from the incident beam. As shown in Fig.~\ref{fig:real_exp} (f), our system achieves a good linear relationship between the relative phase measurements and the beam incident angle, especially after averaging to remove distortions induced by hardware errors. Considering the errors in manually tuning the gimbal rotation, our result demonstrates the accuracy and stability of the proposed reference-free phase sensors when measuring the free-space light field.

\subsection{Application1: phase imaging}
The most natural example proposed application of the phase sensor array is in microscopic phase imaging. 
Phase unwrapping is an important technique that solves the $2\pi$ phase ambiguity in the measured phase (wrapped between $[0, 2\pi]$). In an ideal system without measurement errors (noise), this process can be accurately resolved based on simple differentiation. 
However, small errors in the raw measurements are exaggerated in the unwrapping process~\cite{phasenet, phasenetv2}. As an example, we simulate a $128\times128$ resolution scene in Fig.~\ref{fig:application} (a), with $\textrm{SNR}=20$dB. Both the 2D ``split chain'' sensor array and the 2D grid sensor array achieves close to the ground-truth wrapped phase image. We use a phase unwrapping library based on the noncontinuous path algorithm~\cite{phase_unwrap}. The small errors in the 2D ``split chain'' measurement result leads to large deviations from ground truth in the unwrapped phase images. On the other hand, the phase measurement of the redundant 2D grid design remains robust in the unwrapping process.

\subsection{Application2: on-chip phase monitoring}
\label{sec:app1}
Apart from measuring optical phase of a free-space light field, the proposed system can also be applied in on-chip phase monitoring. Here we show one simulated example in an optical phased array (OPA)~\cite{opa_review}]. OPAs are widely applied in communication~\cite{opa_comm1, opa_comm2}, remote sensing~\cite{3d1, opa_lidar1, opa_lidar2}, and augmented reality~\cite{opa_ar}. An OPA consists of multiple light-emitting antennas, and by adjusting the phase and amplitude in each antenna, an optical power distribution (pattern) is projected to the far field. Accurate phase monitoring is required to generate the desired far field pattern. We show an example in Fig.~\ref{fig:application} (b). The phase monitoring is here proposed to be implemented with a dual-layer photonic circuit~\cite{dual_layer} for simpler waveguide routing. Small amount of power is tapped out from the first layer into the second layer for phase monitoring, and we use black arrows to visualize the optical power application. Previously, chain configuration phase sensor arrays have been demonstrated for this application, in a relatively small OPA~\cite{opa_monitor1, opa_monitor2}. When the number of antenna elements increases, error accumulation becomes more significant. We simulate a 1D OPA with $N = 1024$ antennas. We assume low SNR $= 7$dB in each phase sensor, since only a small amount of optical power is tapped out for phase monitoring to keep the optical efficiency of OPA high. As shown in Fig.~\ref{fig:application} (c), with more robust phase sensor array configurations (``split chain'', ``bisection''), both the phase profile and the far field patterns have significantly higher quality compared to the naive chain configuration.

\section{Discussions}
\label{sec:self-config}
\begin{figure}[t]
    \includegraphics[width=0.85\linewidth]{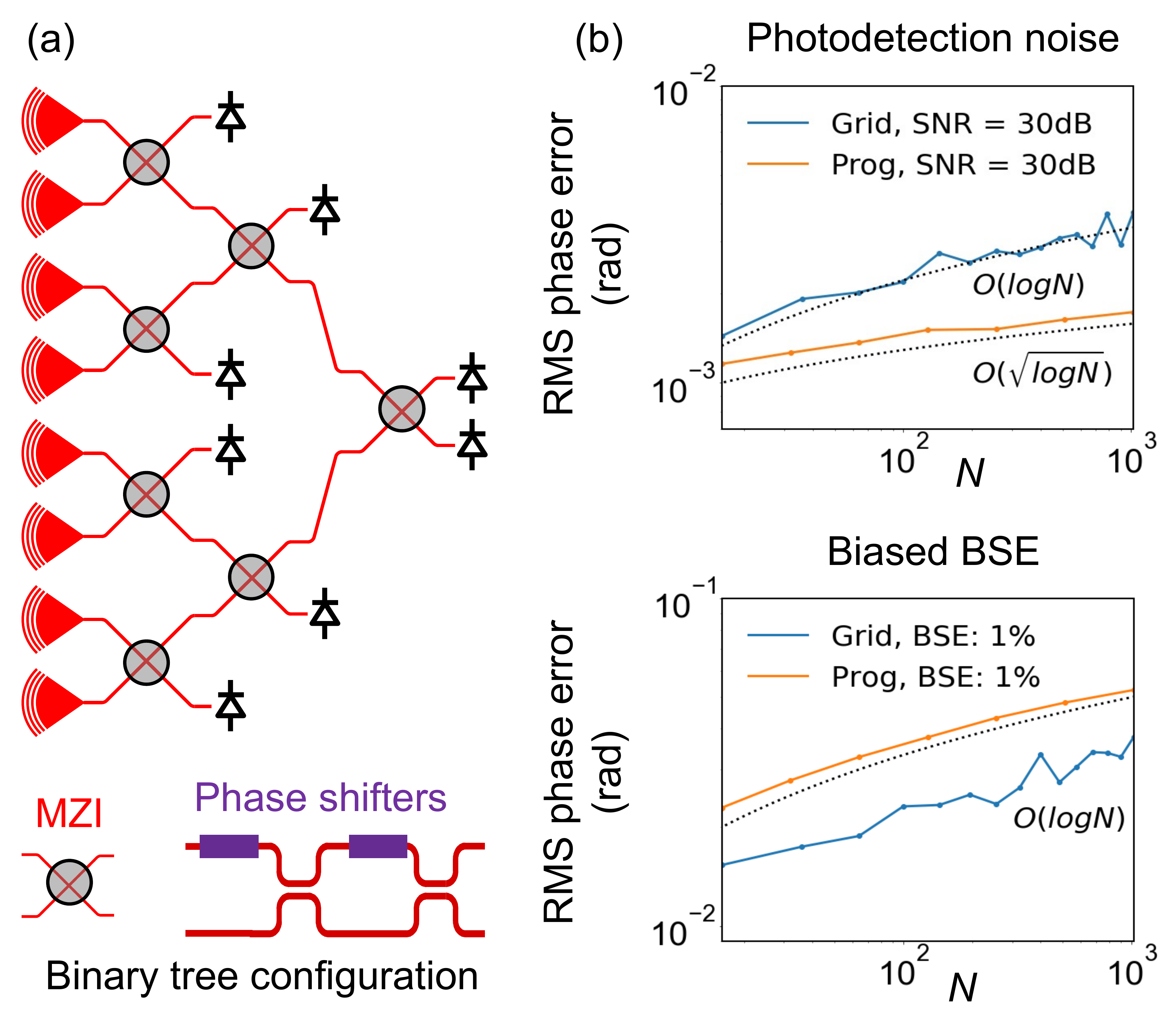}
    \caption{Comparison between progressive approach~\cite{dab_selfconfig} and proposed phase sensor array approach. (a) Schematic showing the photonics circuit of the progressive phase measurement approach~\cite{dab_selfconfig} with a binary tree architecture. (b) Scalability comparisons between progressive and proposed phase sensor array (grid configuration).}
 \label{fig:self-config}
\end{figure}

Here we perform similar phase sensing accuracy analysis on the progressive self-configuration approach~\cite{dab_selfconfig} (the most relevant prior art) and summarize the comparisons with the proposed approach in Fig.~\ref{fig:self-config} and Table.~\ref{tab:self-config}. A schematic of the architecture used in progressive approach is shown in Fig.~\ref{fig:self-config} (a). The circuit consists of $\textrm{log}N$ layers of MZI interferometers, each containing two tunable phase shifters and two $50/50$ beam spitters. 

We suppose the photodetector exposure time is set to $t_0$ to achieve a specific SNR, and the phase shifter tuning time is $t_1$ due to the electronic/thermal bandwidth limit. The proposed low-latency phase sensor requires $2(t_0+t_1)$ (active) or $t_0$ (passive) measurement time. In the progressive approach, the phase measurement process is conducted by nullifying optical powers in each circuit layer progressively~\cite{dab_selfconfig}. A single nullification process can be implemented with a phase shifter sweeping + local feedback control~\cite{jitter}. Although this process is able to generate accurate measurement results, it involves at least tens or hundreds of phase shifter tuning and photodetector exposure periods. To make a fair comparison, we use five measurement steps for each nullification process. Also, since optical power accumulates after each nullification step, the exposure time used in the next layer can be reduced accordingly (we assume the exposure time can be arbitrarily short for simplicity. This results in a total time complexity $\sim 2t_0 + 5t_1\textrm{log}N$. When $t_1$ is longer or comparable to $t_0$ (which is typical in current systems), this is a considerable increase in latency compared to the proposed phase sensor array.

Other important metrics include computational cost and electronics complexity. Since the progressive approach relies on tunable MZI interferometers, it is hard to reduce the form factor, especially for the tunable phase shifters, as discussed in Sec.~\ref{sec:single_sensor}. Also, analog control of the phase shift with high enough precision is required. On the other hand, the proposed phase sensor array can be either totally passive or with a binary phase shifter control (switching between $0$ and $\pi/2$). The progressive approach requires one optical field back-propagation simulation, which takes $O(N)$ computational cost, while the proposed phase sensor array requires one matrix-vector multiplication, which takes $O(N^2)$ computational cost. This is a comparatively minor drawback since computation in the digital domain is comparatively cheaper.

As shown in Fig.~\ref{fig:self-config} (b), when only photodetector noise is present, scalability of the progressive approach can achieve $O(\sqrt{\textrm{log}N})$, better than the designed configurations. While with biased hardware error, the scalability is comparable to the proposed phase array (both $O(\textrm{log}N)$) but with a slightly higher coefficient. The good scalability of the progressive approach is due to the short error accumulation path length in the binary tree architecture (similar to the ``bisection'' configuration in Fig.~\ref{fig:configs}), and to the optical power accumulation in the progressive programming process. Concentrated optical power leads to high signal-noise-ratio and low phase sensing errors in each element. We discuss applying the low-latency, passive phase sensors in such binary tree architecture in more detail in the supplementary material. Although the path length remains short, without progressive optical power accumulation, such a system does not possess satisfying accuracy and scalability.

Apart from the hardware error and measurement noise analysis, there is another advantage of having redundant elements in the phase sensor array. As an example, suppose the right-most MZI node in Fig.~\ref{fig:self-config}(a) is malfunctioning, the entire measurement fails while there is no method to easily tell which part of the circuit is causing the error. However, a redundant phase sensor array (as proposed in Fig.~\ref{fig:configs}) is not influenced by a few malfunctioning/broken elements. Also, from the least-square solving process (Eqn.~\ref{eqn:least_square}), it is easy to distinguish malfunctioning elements with the residual error. We show in supplementary material that such a system is still robust with $>10\%$ malfunctioning/broken elements. 

\begin{table}[t]
\begin{tabular}[b]{|c|c|c|}\hline
   & Progressive~\cite{dab_selfconfig} & Phase sensor array \\ \hline
  
  Redundant & No & Yes \\
  \hline
  \thead{Time complexity} & \textcolor{red}{$8t_0 + 5t_1\textrm{log}N$} & \makecell{\textcolor{teal}{$2(t_0 + t_1)$ or $t_0$}} \\
  \hline
  Scalability & \textcolor{teal}{$O(\textrm{log}N)$}  & \textcolor{teal}{$O(\textrm{log}N)$} \\
  \hline
  \thead{Computational cost} & \textcolor{teal}{$O(N)$} & \textcolor{red}{$O(N^2)$} \\
  \hline
  \thead{Electronics \\ complexity} & \textcolor{red}{DAC required} & \makecell{\textcolor{teal}{Digital only}\\\textcolor{teal}{or passive}}\\
  \hline
\end{tabular}
\caption{Summary of metrics. Red color indicates worse, green color indicates better.}
 \label{tab:self-config}
\end{table}

\section{Conclusion}
\label{sec:conclusion}
 Fast and simple optical phase measurement is important in optical communication and sensing. In this paper, we propose a low-latency, reference-free optical phase sensor array based on integrated photonics. 
 We demonstrate post-processing algorithms and circuit design rules (e.g., connection pathlength, redundancies, and polarities of phase sensor elements) that enable high accuracy and scalability in the presence of measurement noise and hardware errors.
 This leads to improved accuracy and robustness in applications across disciplines, including microscopy, remote sensing, optical computing, and optical communication.

\section{Acknowledgement}
This work was supported by AFOSR grants FA9550-17-1-0002, FA9550-18-1-0186, and FA9550-21-1-0312.

\bibliography{phasemeasurement}

\appendix
\section{Derivations of Eqn.~\ref{eqn:array_error}}
Following Eqn.~\ref{eqn:least_square}, $\hat{\textbf{p}}$ can be expressed with $M^+$, the pseudo-inverse of measurement matrix $M$. The root mean squared (RMS) phase sensing error $\textrm{RMS}_p$ is defined to be the root mean square difference between the estimated phase profile $\hat{\textbf{p}}$ and the ground truth $\textbf{p}$. 
\begin{eqnarray}
\label{eqn:derive1}
\hat{\textbf{p}} = M^+\textbf{p}_H
\;\;\; M^+ = (M^TM)^{-1}M^T
\\
\nonumber
\mathbb{E}[\textrm{RMS}] = ||\hat{\textbf{p}} - \textbf{p}|| = ||M^+\textbf{e}_H||
\end{eqnarray}

Where $\textbf{e}_H$ is the errors in measurements $\textbf{p}_H$. Note that we always choose the first input port as phase reference and its phase is always manually set to zero. So when computing $\mathbb{E}[\textrm{RMS}]$, the first element of $M^+\textbf{e}_H$ is not included. As discussed in Sec.~\ref{sec:single_sensor}, errors in single phase measurement can be random or biased. We utilize the statistical properties $\mathbb{E}[e_{Hi}e_{Hj}] = \sigma_e^2\delta_{ij}$ ($\delta_{ij}$ is the Kronecker delta) for random measurement error and $\mathbb{E}[e_{Hi}e_{Hj}] = e_0^2$ for biased measurement error in our derivation.

\begin{flalign}
\label{eqn:derive2}
&\mathbb{E}[\textrm{RMS}]^2 = \sum_i\;[\;\sum_j M_{ij}^+e_{Hj}\;]^2&
\\
\nonumber
&= \sum_i(\sum_j M_{ij}^+e_{Hj}) (\sum_k M_{ik}^+e_{Hk})&
\\
\nonumber
&= \sum_{ijk} M_{ij}^+ M_{ik}^+ e_{Hj}e_{Hk}&
\\
\nonumber
&= \left\{ \begin{array}{l}
\textrm{Random:}\; \sigma_e^2\sum_{ijk} M_{ij}^+ M_{ik}^+ \delta_{jk}= \sigma_e^2\sum_{ij}M^+[ij]^2
\\
\textrm{Biased:}\; e_0^2\sum_{ijk} M_{ij}^+ M_{ik}^+ = e_0^2\sum_{ij}(M^{+T}M^+)[ij]
\end{array} 
\right.&
\end{flalign}

This equation establishes a simple relationship between the average phase sensing accuracy and the matrix norm of the measurement matrix $M$.

\begin{figure*}[t]
    \includegraphics[width=0.85\linewidth]{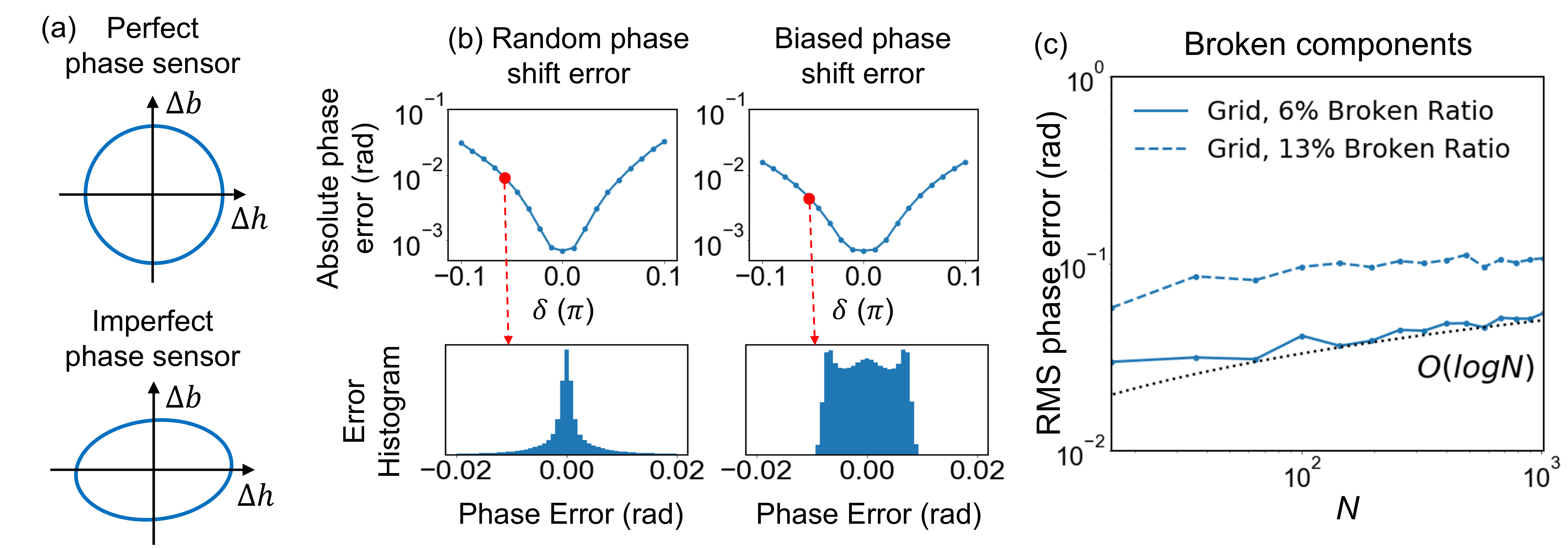}
    \caption{(a) $\Delta h-\Delta b$ curve with perfect (upper) and imperfect (lower) phase sensor when relative phase between two input ports ranges from $0$ to $2\pi$. The goal of calibration is to measure and store the shape of the $\Delta h-\Delta b$ curve (a general ellipse). (b) Average absolute phase error and error histograms with phase shift error, with similar behavior as with beam-splitting ratio error discussed in the main text. Therefore, all conclusions can be straight-forwardly extended to phase shift error case. (c) Phase sensing accuracy with broken/malfunctioning elements. The redundant 2D grid phase sensor array still maintains good scalability with $13\%$ broken elements, which is infeasible in non-redundant phase sensor array or self-configuration systems~\cite{dab_selfconfig}.}
 \label{fig:hardware_error_more}
\end{figure*}

\section{More Discussions on hardware errors}
\subsection{Phase sensor array calibration}
Free-space interferometers encounter hardware errors similar to those discussed in this work for the integrated phase sensor. To mitigate these errors, calibration is performed. Here, we provide a brief summary of the calibration process, and readers can refer to \cite{hetero_calib1, hetero_calib2} for more detailed discussions. As shown in Fig.~\ref{fig:hardware_error_more} (a), we define $\Delta h = h_+ - h_-$ and $\Delta b = b_+ - b_-$. In the case of a perfect phase sensor, when the relative phase of two input ports changes from $0$ to $2\pi$, $\Delta h$ and $\Delta b$ form a perfect circle. However, in the presence of hardware errors, such as beam-splitting ratio error, phase shift error, and photodetector error, the $\Delta h-\Delta b$ curve becomes a general ellipse. The objective of calibration is to measure at least six points (more is better) on the ellipse to determine its shape, which is represented as an affine transformation matrix. This matrix can be used to correct errors at each single phase sensor during operations.

\subsection{Phase shift error}
Figure~\ref{fig:hardware_error_more} (b) presents the measurement error of a single phase sensor under biased or random phase shift error (we assume that the $\pi/2$ phase shift is actually $\pi/2 + \delta$). We use the same simulation settings as in Fig.~\ref{fig:single_sensor_error} of the main text. The average absolute phase error and error histograms in this scenario are similar to those observed with beam-splitting ratio error. Hence, all conclusions drawn from the analysis of beam-splitting ratio error can be directly extended to handling phase shift error.


\subsection{Broken array elements}
In the main text, we assume that each phase sensor element in the circuit is identical. However, certain elements may malfunction or certain input ports may receive very low signal levels, violating this assumption. Mathematically, these situations are abstracted as certain edges in the circuit graph being broken, as shown in Fig.\ref{fig:hardware_error_more} (c). If the circuit is non-redundant, then the full phase profile measurement fails. However, the more redundancies in the circuit, the more robust it is to this issue. To demonstrate this, we simulate the performance of the "grid" phase sensor array with $6.25\%$ and $12.5\%$ broken elements, as well as $1\%$ beam splitter error and SNR $= 30$dB photodetection noise in the simulations. As shown in Fig.\ref{fig:hardware_error_more} (c), the system remains robust under these extreme conditions.

\section{Polarization state sensing}
\begin{figure}[t]
    \includegraphics[width=0.5\linewidth]{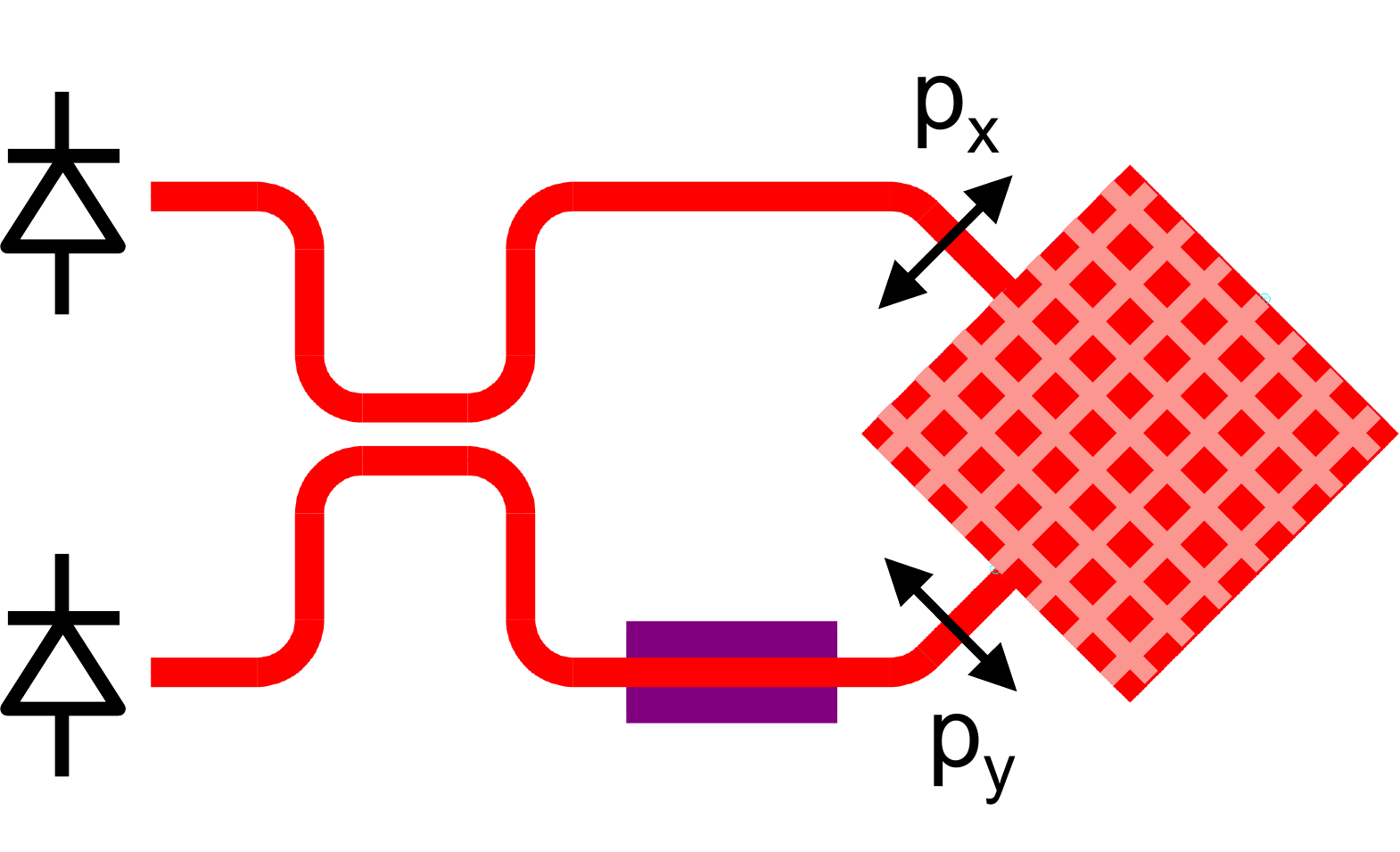}
    \caption{Schematic showing polarization state sensing with the proposed phase sensor.}
 \label{fig:polarize}
\end{figure}

The proposed system measures relative phases between the optical fields in multiple single-mode waveguides. As discussed in ~\cite{dab_selfconfig}, these optical modes can be converted from free space optical fields by an optical interface. In the main text, we use conventional grating couplers as this interface. Another example is to utilize polarization sensitive grating couplers~\cite{polarize_grating_coupler}. As shown in Fig.~\ref{fig:polarize}, different polarization components in the incident light field is split into two single-mode waveguides. With the phase sensor, relative phase between the two polarization components can be measured.

\section{Hardware implementation}
\begin{figure*}[t]
    \includegraphics[width=0.7\linewidth]{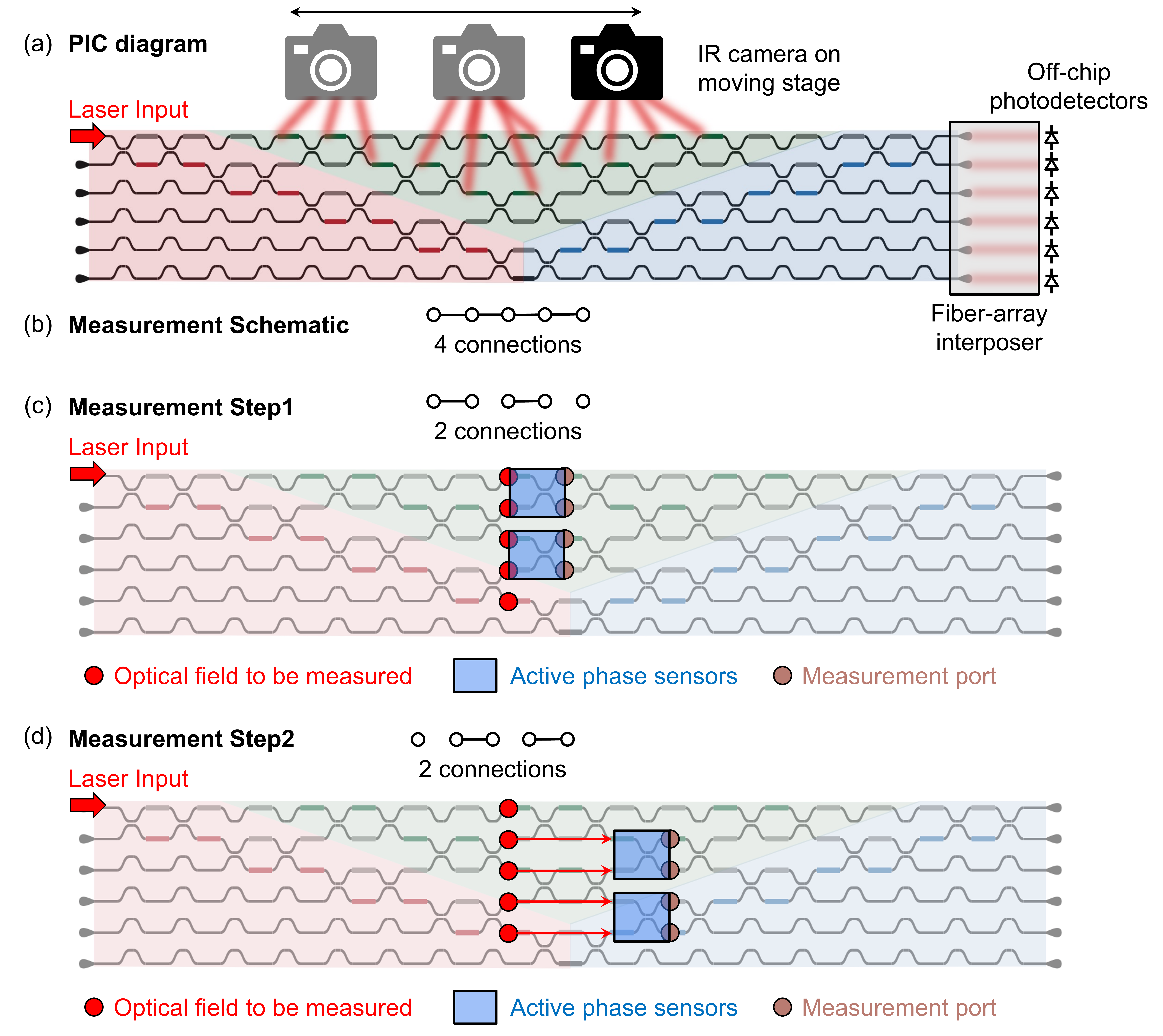}
    \caption{Circuit diagram and the hardware implementation of on-chip phase sensing with a general-purposed PIC mesh~\cite{sunil_insitu}.}
 \label{fig:hardware_implement}
\end{figure*}

We use the same photonic integrated circuit (PIC) platform as~\cite{sunil_insitu}. The PIC diagram and our hardware implementation are shown in Fig.~\ref{fig:hardware_implement} (a). The PIC is a $6\times6$ triangular mesh~\cite{sunil1}. Using single-mode fiber, we input laser power into one of the input ports and detect the optical power in the waveguides with the grating coupler taps distributed in the whole mesh and a moving IR camera (for details, please refer to~\cite{sunil_insitu}). We show the phase measuring experiment process in Fig.~\ref{fig:hardware_implement}(c),(d). We program the mesh to generate desired optical fields in $5$ waveguides (denoted as red dots). Then we divide these ports into pairs and program part of the mesh (blue bounding boxes) as active phase sensors to measure the relative phase. We use multiple measurement steps to measure all desired pairs of ports. In Fig.~\ref{fig:hardware_implement}, we show the full measurement process (two steps) to measure the phase profile with 4 connections, while it is straightforward to measure phases between more pairs (e.g., the 6 connection and 8 connection configurations in the main text, Fig.~\ref{fig:real_exp}). In measurement step$2$, we program the mesh to route the $4$ lower optical fields (red arrows) to another layer in the mesh. Note that to make fair comparisons between different configurations, we control the input laser power and optical field generation to emulate the optical power distribution in different configurations with a fixed total amount of input power.

\section{$2\pi$ ambiguity in phase measurement}
This difference prohibits utilizing native least-square solvers (e.g., with QR matrix decomposition~\cite{linalg_textbook}). Similar problems are encountered in 3D reconstruction from surface normal field~\cite{ramesh_grad_field, grad_field_1}, and the rigorous mathematical solution involves a or treating the problem as a Poisson equation with known boundary conditions. An easier alternative is to use a non-redundant part (a path that connects all input ports without redundancy) within the redundant phase sensor array to initialize the phase profile estimation, the rough range of phase at each input port is determined with this initialization. The least-square method then refines the phase estimation with all phase measurements in the redundant array.

\section{Relationship to Shack-Hartmann sensor}
Here we briefly discuss the relationship between the proposed technique and the conventional Shack-Hartmann wavefront sensor (SHWS). Generally, both systems measure the phase gradients. However, in SHWS this gradient is encoded in the spatial displacement, which leads to the trade-off between spatial and phase resolutions. In our system this gradient is directly measured with the I/Q detection, and is thus more accurate and supports higher spatial resolution and scalability. Also, PIC allows more flexible phase gradient measurements between non-adjacent neighbors (e.g. ``Grid+Diag'' configuration in Fig.~\ref{fig:configs} (a)). With this redundancy, the overall sensing robustness is further enhanced. On the other hand, the major advantage of SHWS is that it can be easily applied to wide wavelength range or even white light, while the optical bandwidth of PIC phase sensor is limited by the photonics building-blocks. However, with broad-band design of photonics components (e.g., grating couplers~\cite{grating_coupler_wb}, beam splitters~\cite{beam_splitter_wb}) and calibration (detailed in the supplementary material), a wider wavelength range can be achieved.

\begin{figure*}
    \centering
    \includegraphics[width=\linewidth]{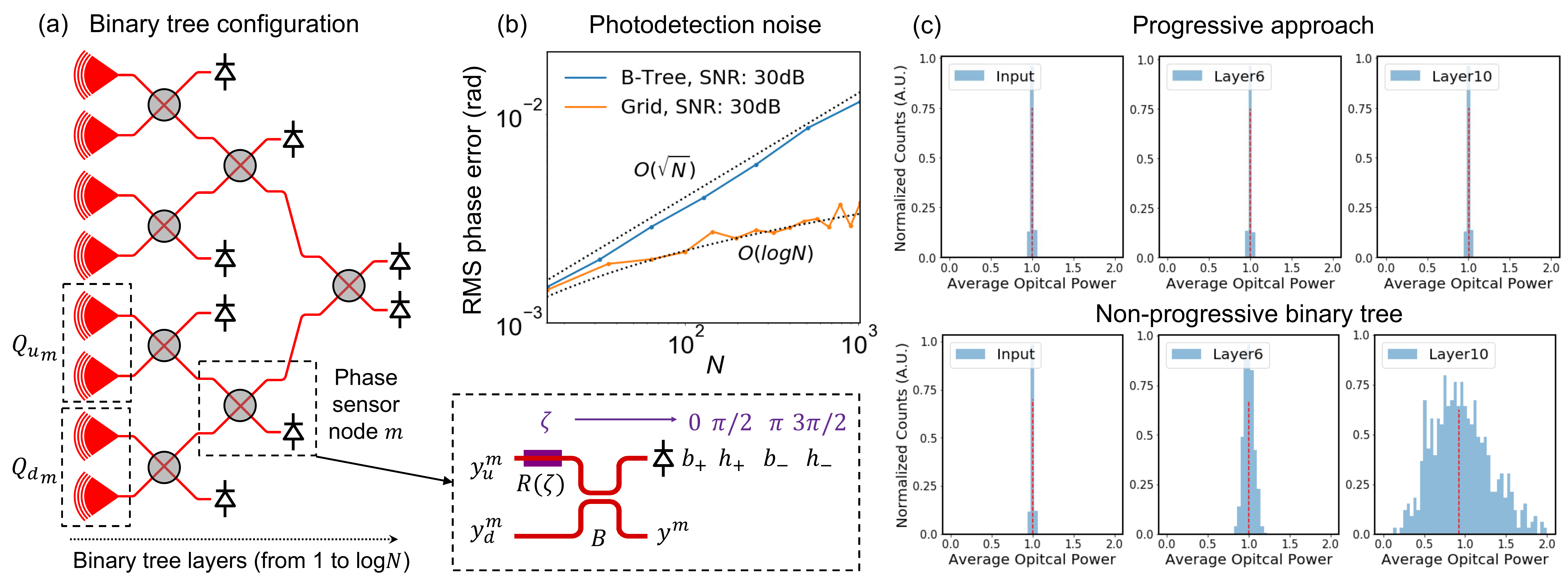}
    \caption{Non-progressive phase measurement with a binary tree architecture. (a) Photonic circuit configuration, each node is an active or passive phase sensor instead of an MZI. (b) Root-mean-squared (RMS) phase sensing error in presence of photodetection noise. Non-progressive binary tree configuration has $O(\sqrt{N})$ error scaling despite short error accumulation paths. (c) This unfavorable error scaling is attributed to the loss of optical power in propagation, different from that in the progressive approach.}
    \label{fig:multiport}
\end{figure*}

\section{Non-progressive phase sensing with binary tree architecture}
As discussed in main text, Sec.~\ref{sec:self-config}, we analyzed the scalability of the progressive self-configuration approach. It turns out to have high robustness due to the short error propagation chain in the binary tree architecture and the power accumulation in the progressively programming process. Therefore, a natural extension is to integrate the single-shot phase sensors into such architecture. However, as we show in the following discussions, such combination does not provide satisfying robustness, due to the missing optical power accumulation capability.

As shown in Fig.~\ref{fig:multiport} (a), the non-progressive binary tree phase sensor array has same high-level architecture as in the progressive approach. However, each node is a active or passive phase sensor element, instead of an MZI. During operation, the phase sensor tree takes one measurement if the nodes are passive or four measurements if the nodes are active (since there is only one photodetector connected to the node output, we set the phase shift to $0, \pi/2, \pi, 3\pi/2$ for I/Q detection). 

Phase of the input light field is estimated with a recursive algorithm from the measurement results. We denote the global phase at $n$th input port $\theta_n$. At the start of the recursive algorithm, all $\theta_n$ as initialized to $0$. During the recursion process, at phase sensor node $m$ with upper port $u_m$, lower port $d_m$, and phase measurement results $b_{\pm,m}, h_{\pm,m}$, the local relative phase $\gamma_m$ is calculated as in the main text
\begin{equation} \label{eqn:localphase}
\gamma_m = \textrm{arctan}((b_{+,m} - b_{-,m})/(h_{+,m} - h_{-,m}))
\end{equation}

This local relative phase contributes to a global phase difference between all input ports within the sub-tree with the upper port as root (denoted by $Q_{u_m}$ as in Fig.~\ref{fig:multiport}) and all input ports within the sub-tree with the lower port as root (denoted by $Q_{d_m}$ as in Fig.~\ref{fig:multiport}). More concretely, given the input fields $y_{u_m}, y_{d_m}$ calculated at a deeper recursion step, this global phase difference can be calculated as

\begin{equation} \label{eqn:nodephase}
    \begin{aligned}
        \phi_m &= \gamma_m - \arg\left(\frac{y_{u_m}}{y_{d_m}}\right)
    \end{aligned}
\end{equation}

where $\textrm{arg}()$ is the angle of the complex number. For all nodes in $Q_{u_m}$, we add their global phase $\theta_n$ by $\phi_m$.
We further compute the output field $y_m$ as 
\begin{equation} \label{eqn:field}
    y_m = \frac{1}{\sqrt{2}} \left(e^{i \phi_m} y_{u_m} + y_{d_m} \right).
\end{equation}

This output field is passed to a shallower recursion step at phase sensor node $m'$ and serve as the input field $y_{u_{m'}}$ or $y_{d_{m'}}$. At the deepest level of recursion, which corresponds to the input phase sensors, $Q_{u_m}$ and $Q_{d_m}$ only contains a single input port, and $y_{u_m}, y_{d_m}$ are set as the input field magnitudes $|x_{u_m}|, |x_{d_m}|$. The whole recursion process is summarized in Alg. \ref{alg:phasecalculation}.

We conducted a simulation of the non-progressive binary tree architecture using the same random input field settings as detailed in the main text. The results, illustrated in Fig.\ref{fig:multiport} (b), indicate that even with a short error accumulation path ($O(\sqrt{\textrm{log}N})$), the phase sensing error scales with $O(\sqrt{N})$ in the presence of photodetection noise. This is attributed to optical power loss during propagation. For simplicity, we assign each layer (column) in the binary tree architecture a layer index and denote layers closer to the tree root to be ``higher''. As shown in Fig.~\ref{fig:multiport} (c) upper panel, for the progressive approach\footnote{We assume shorter exposure time for higher layers, as discussed in main text, Sec.~\ref{sec:self-config}}, average optical power does not change when light propagates from the first layer to higher layers. Note that we compute the average optical power among all the MZI nodes at the same layer. This is because at each layer the MZI are set to concentrate all optical power into a single output port. This optical power accumulation capability, along with the exponentially decaying exposure time, guarantees the constant average optical power during propagation independent on the incident optical field. On the other hand, in a non-progressive binary tree consists of active phase sensors, average optical power has large variance at higher layers, as shown in Fig.~\ref{fig:multiport} (c) lower panel. This indicates that average optical power at higher layers can be very low with some incident optical field, leading to low signal-noise-ratio (SNR) and big errors in the phase measurements. What's more, since phase measurement results at higher layers influences $\theta_n$ at more input ports (there are more input ports in $Q_{u_m}$ when node $m$ is at higher layer), this inaccurate phase measurement leads to big errors in the estimated phase profile.

\begin{algorithm}[H]
    \caption{Phase calculation}
    \label{alg:phasecalculation}
    \begin{algorithmic}[1]
        \Function{UpdateTheta}{$|\boldsymbol{x}|$, $|\boldsymbol{h}|$, $|\boldsymbol{b}|$, $\mathcal{M}$, $m$, $\boldsymbol{\theta}$}
            \State Get $Q_{u_m}, Q_{d_m}$ from $\mathcal{M}$.
            \If{$|Q_{u_m}| > 1$}
                \State $y_{u_m} \gets$ \textsc{UpdateTheta}($|\boldsymbol{x}|$, $|\boldsymbol{h}|$, $|\boldsymbol{b}|$, $\mathcal{M}$, $u_m$, $\boldsymbol{\theta}$)
            \Else
                \State $y_{u_m} \gets |x|_{u_m}$
            \EndIf
            \If{$|Q_{d_m}| > 1$}
                \State $y_{d_m} \gets$ \textsc{UpdateTheta}($|\boldsymbol{x}|$, $|\boldsymbol{h}|$, $|\boldsymbol{b}|$, $\mathcal{M}$, $d_m$, $\boldsymbol{\theta}$)
            \Else
                \State $y_{d_m} \gets |x|_{d_m}$
            \EndIf
            \State $\Delta h \gets |h_{+, m}|^2 - |h_{-, m}|^2$
            \State $\Delta b \gets |b_{+, m}|^2 - |b_{-, m}|^2$
            \State $\gamma_m \gets \gamma(\Delta h, \Delta b)$ \Comment Eq. \ref{eqn:localphase}
            \For{$n \in Q_{u_m}$}
                \State $\theta_n \gets \theta_n + \phi(\gamma_m, y_{u_m}, y_{d_m})$  \label{algline:updateloop} \Comment Eqs. \ref{eqn:nodephase}
            \EndFor
            \State \Return $y_m(\gamma_m, y_{u_m}, y_{d_m})$ \Comment Eqs. \ref{eqn:field}
        \EndFunction
        \item[]
        \Function{GetPhases}{$|\boldsymbol{x}|$, $|\boldsymbol{h}|$, $|\boldsymbol{b}|$, $\mathcal{M}$}
        \State $\boldsymbol{\theta} \gets \boldsymbol{0}^{N}$
        \State \textsc{UpdateTheta}($|\boldsymbol{x}|, |\boldsymbol{h}|, |\boldsymbol{b}|, \mathcal{M}, 1, \boldsymbol{\theta}$)
        \State \Return $\boldsymbol{\theta}$
        \EndFunction
    \end{algorithmic}
\end{algorithm}

\end{document}